\documentclass[12pt]{article}
\usepackage[left=2cm,right=1.5cm,
    top=2cm,bottom=2cm,bindingoffset=0cm]{geometry}
\usepackage{graphicx,times}
\usepackage{amssymb,amsmath}
\usepackage{color}
\usepackage{floatrow}
\usepackage[cp1251]{inputenc}
\usepackage[T1,T2A]{fontenc}
\usepackage[english]{babel}
\usepackage{natbib}

\begin{document}
\begin{figure}
    \centering
    \includegraphics[width=0.9\textwidth]{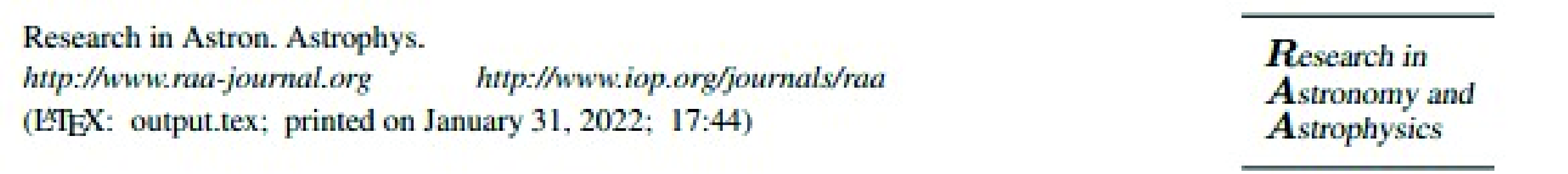}
\end{figure}
\begin{center}
\begin{Large}

\textbf{New features of the pulsar B0950+08 radiation \\ at the frequency of 111 MHz}

\end{Large}

\vspace{5mm}
V. M. Malofeev, I. F. Malov, O. I. Malov, D. A. Teplykh \\
\vspace{5mm}

\it{P.N.Lebedev Physical Institute of the Russian Academy of Sciences, \\
Leninskii pr. 53, Moscow, 119991, Russia;\\ teplykh@prao.ru}

\vspace{5mm}
Accept 13-Dec-2021
\end{center}
\abstract{Results of long time observations of the pulsar B0950+08 are given. These observations were carried out at the LPA radio telescope at the frequency of 111 MHz from January of 2016 to May of 2019 (450 days). A strong variability in emission of this pulsar has been detected with changes of signal to noise ratios hundreds of times. Part of the long-time flux density variability can be explained by refractive scintillations in the interstellar medium. The existence of radiation between the interpulse (IP) and main pulse (MP)  was confirmed. It was more powerful than at high frequencies. We detected the unusual interpulse and precursor (Pr) radiation on August 1, 2017. On the base of strong 65 interpulses we found the correlations between energies of IP and Pr and between the phase of IP and the distance Pr-IP. It is shown that the observed peculiarities of this pulsar can be explained in the frame of the aligned rotator model. We estimated distances of radiation levels from the center of the neutron star. The calculated value of the initial period of 0.2 sec means that not all pulsars are born with millisecond periods. The large age of the pulsar (6.8 millions of years) and the small angle between its  magnetic moment and the rotation axis (less than 20$^{\circ}$) confirm the suggestion on the pulsar evolution to an alignment. 

keywords: pulsars: individual B0950+08
}

\section{Introduction}   

Most of the known 3000-plus radio pulsars 
(\citeauthor{MHT05}, \citeyear{MHT05}) are characterized by an average pulse width of order of several percents of the period. However, there are pulsars emitting in the wider longitude range. This is specific particularly for radiation at low frequencies. In this case emission is generated at large distances from the surface of the neutron star where the emission cone expands markedly. If the angle between the magnetic moment and the rotation axis is small then the line of sight can be inside the emission cone for a long time. We believe that such a case takes place for the pulsar B0950+08 emitting sometimes during the whole period. For example, such observations were made at 408 MHZ (\citeauthor{HC81}, \citeyear{HC81}  and \citeauthor{PL85}, \citeyear{PL85}) and at 111 MHz (\citeauthor{S12}, \citeyear{S12}). Some features of this pulsar are discussed in our paper. In addition to the main pulse (MP) and interpulse (IP), this pulsar has “the bridge” and “the precursor” located between MP and IP (see, for example, \citeauthor{HC81}, \citeyear{HC81}). They analysed the complex time-frequency structure both of integrated pulses and of individual ones in this pulsar. The frequency dependences of the width of MP and IP as well the distances between IP and MP were presented also. The discussion about the arising of MP and IP from opposite magnetic poles or from a single magnetic pole (the aligned rotator) takes place in a number of papers (see, for example, \citeauthor{PL85}, \citeyear{PL85} and \citeauthor{HC81}, \citeyear{HC81}).

The LPA antenna was upgraded in 2014 and its sensitivity became two times higher. As the result the time-frequency resolution was improved significantly and about 60 new pulsars and RRATs were discovered (\citeauthor{TTO16}, \citeyear{TTO16}, \citeauthor{TTK17}, \citeyear{TTK17}, \citeauthor{TTM18}, \citeyear{TTM18}, \citeauthor{TKT20}, \citeyear{TKT20}). There was the possibility to carry out new observations of the pulsar B0950+08 and detect some interesting features in its radiation. Further we describe the results of these new observations.

The reminder of the paper is organized as follows. In Section 2 we discuss some details of our observations. Section 3 contains the results and their analysis. In Section 4 the interpretation of these results and the discussion are given. In Section 5 the conclusions are summarized.

\section{Observations}

All observations were carried out using the meridional radio telescope LPA (Large Phase Array). Its antenna is the phased array composed of 16384 dipoles. The geometric area of this antenna is more than 70 000 $m^{2}$ and the effective area is   ${47000 \pm 2500\,m^2}$ (\citeauthor{TTO16}, \citeyear{TTO16}). Antenna has 128 space beams with the size of one beam $0.9^{\circ} \times 0.5^{\circ}$. The duration of an observing session is $3.5/cos\,\delta$ minutes, where $\delta$ is the declination. The measurement series lasting about 10-20 days each were carried out from 2016 to 2019. For the processing 460 channels of the 512 channel digital receiver are used. The single channel bandwidth is 4.88 kHz, and the time resolution is 1.23 msec, the total bandwidth is 2.245 MHz. All data is stored in the server. For the processing the special program has been worked out (\citeauthor{MTL12}, \citeyear{MTL12}). We have rather good RFI (radio frequency interference) situation on our radio telescope, especially during night time. We have the dynamic spectrum of every  pulsar period and can detected a pulse RFI in the frequency domain and in the time domain. There are two regular radio stations in our frequency band (2.245 MHz). They occupy 2-3 frequency channels (the band  is equaled to 10-15 kHz). Sometimes we see sporadic pulse RFI which occupy a part of our frequency band or even the entire band. To clean the data, we removed from the reduction such frequency channels. Our estimations show, that the fraction of  RFI data occupy no more than 2\% of the observation time. Here we present the analysis of integrated profiles and single pulses for the pulsar B0950+08. The focus is on the unique event that took place on August 1, 2017.
 
\section{Results and analysis}

\begin{figure}[t]
\centering
	\includegraphics[width=0.9\textwidth]{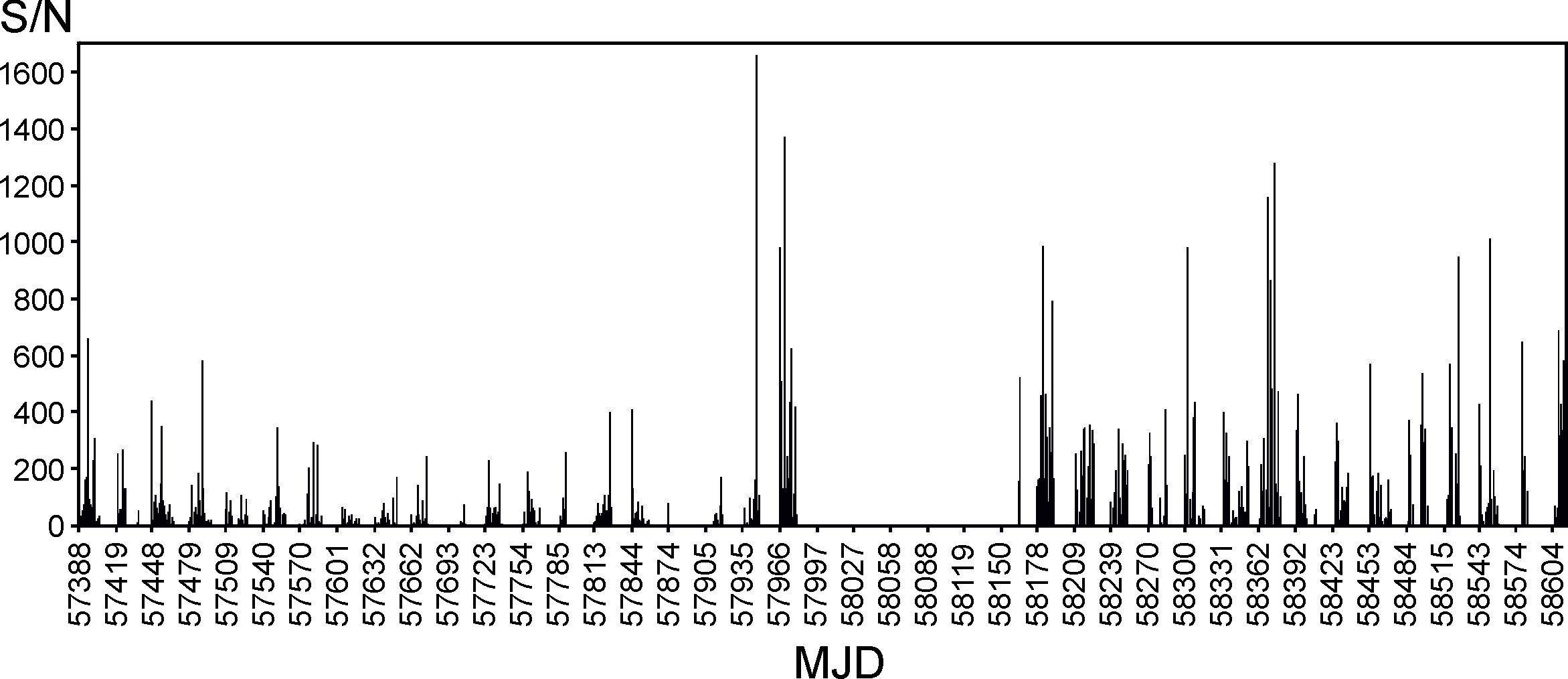}
    \caption{\small S/N variation of emission during 3.5 years of observations.}
\end{figure}
Figure 1 shows the dependence of the signal to noise ratio (S/N) in integrated pulses on the date of observations from January of 2016  to May of 2019 (450 days). In our case, the value S/N is the ratio of the pulse amplitude to sigma of the noise. To calculate $\sigma$ we used the time interval with minimum radiation in the integrated profile of pulsar. This  interval is shown in Fig. 4b. 
It is evident that S/N or flux densities of integrated pulses are extremely variable. Integrated pulses were constructed on the base of 3.2 minutes of observations (764 pulsar periods). These pulses change the intensities on the time scales from one to five days and show power bursts when S/N grows up to the values of 2000 in individual pulses. There are also long quasi-periodic variations of flux densities with scales of several months.

We detected the unusual changes of the flux densities of this pulsar on August 1, 2017. These changes took place on both millisecond scales inside the period (Fig. 2) and second-minute scales (Fig. 3) for the observation session duration of 3.2 min. Most likely these changes are caused by the radiation mechanism itself, however, some of them are possibly due to scintillations caused by inhomogeneities of the interstellar plasma.

\subsection{Influence of interstellar scintillations and polarization}

There are two types of interstellar scintillations, namely, diffractive (DISS) and refractive (RISS) ones. The first ones are characterized by short periods, the second ones by long times. \citeauthor{SMP95} (\citeyear{SMP95}) have given the empirical expressions for estimations of a decorrelation bandwidth ($\Delta\nu_{d}$), a decorrelation time ($\tau_{d}$) and a scintillation index ($m$) describing diffractive scintillations. These expressions have been obtained on the base of the large number of measurements of the parameters mentioned at the number of frequencies for dozens of pulsars located at different distances or different dispersion measures, see the references in \citeauthor{SMP95} (\citeyear{SMP95}). We will use the following equalities:

\begin{equation}
    \Delta\nu_{d}(\text{kHz})=8.8\left(\frac{\nu}{\nu_{0}}\right)^{4}\left(\frac{DM}{DM_{0}}\right)^{-1.45}\,\text{for}\,DM\leq 20\,pc\cdot cm^{-3} 
\end{equation}
and
\begin{equation}
    \tau_{d}(\text{min})=10\left(\frac{\nu}{\nu_{0}}\right)^{1.1}\left(\frac{DM}{DM_{0}}\right)^{-0.4}\text{for}\,DM\leq 30\,pc\cdot cm^{-3}  ,
\end{equation}
where  $\nu_{0}=1$\,GHz,  $DM_{0} =30\,pc\cdot cm^{-3}$. Using $DM = 2.97\,pc\cdot cm^{-3}$ from \citeauthor{MHT05} (\citeyear{MHT05}) for the pulsar B0950+08  we have $\Delta\nu_{d}$ = 38\,kHz and $\tau_{d}$ = 2.2\,min at the frequency of $\nu$ = 111\,MHz. \citeauthor{S06} (\citeyear{S06}) has given $\Delta\nu_{d}$ = 200\,kHz and   $\tau_{d} >$ 3.3\,min. Using results of measurements at frequency of 60 MHz  (\citeauthor{KWR20}, \citeyear{KWR20}) and the value of  $\Delta\nu_{d}=32$\,kHz at this frequency we obtain 480 kHz for this quantity at our frequency on the base of equations (1) and (2). Measurements of \citeauthor{BMJ16} (\citeyear{BMJ16}) at 154 MHz give $\Delta\nu_{d}$ =4.1\,MHz  and $\tau_{d}$ =28.8\,min.  Then we have  $\Delta\nu_{d}$ =350\,kHz, $\tau_{d} = 21$ min at 111 MHz.  Almost the same results are obtained, if we use the dependences from (\citeauthor{C86}, \citeyear{C86}): $\Delta\nu_{d} = (\nu/\nu_{0})^{4}$, $\tau_{d}=(\nu/\nu_{0})^{1.2}$. The scintillation index decreases markedly due to the smoothing of scintillations with the decorrelation bandwidth $\sim$\,38\,kHz inside of our receiver bandwidth (2.3\,MHz). As was shown by \citeauthor{S71} (\citeyear{S71}) the scintillation index decreases to 0.25 for the ratio of these two bands equaled to 60. Hence, diffractive scintillations can not explain not only millisecond and second intensity variations but minute ones as well. Let us estimate an influence of refractive scintillations. For this process \citeauthor{SMP95} (\citeyear{SMP95}) obtained formulas for the scintillation index ($m_{r}$) and the decorrelation time ($\tau_{r}$) on the base of long time measurements by \citeauthor{GRC93} (\citeyear{GRC93}) at the frequency of 74 MHz:

\begin{equation}
    m_{r}=0.6\left(\frac{\nu}{\nu_{0}}\right)^{1/2}\left(\frac{DM}{DM_{0}}\right)^{-1/3} 
\end{equation}
and
\begin{equation}
    \tau_{r}(\text{days})=40\left(\frac{\nu}{\nu_{0}}\right)^{-2}\left(\frac{DM}{DM_{0}}\right)^{1.5}  
\end{equation}

\begin{figure}[p]
\centering
	\includegraphics[width=0.8\textwidth]{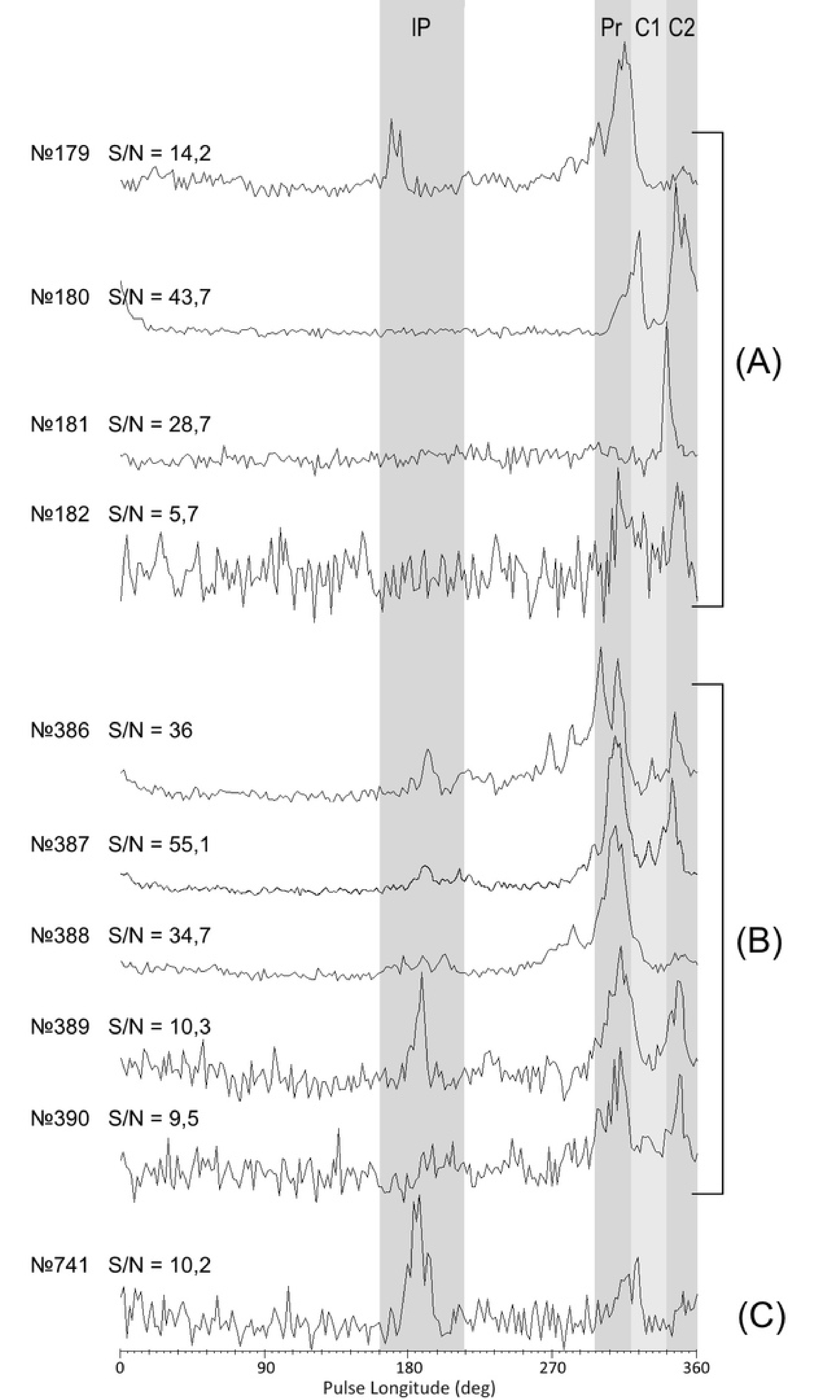}
    \caption{\small Examples of successive individual pulses on August 1, 2017.: changes of intensity for all components from one pulse to another (A),  relative stability of radiation during several periods (B) and  the case when IP stronger than MP (C). The numbers of individual pulses are shown.}
\end{figure}

As the result we have $m_{r}=0.42$, $\tau_{r} = 101$\,days. Let us compare these values with the results of measurements at fairly close frequencies. \citeauthor{GRC93} (\citeyear{GRC93}) have given $m_{r} = 0.45$, and $\tau_{r} = 3^{d}.4$ at 74 MHz. Conversion to frequency of 111 MHz using equations (3) and (4) gives $m_{r} = 0.55$ , and $\tau_{r} = 1^{d}.5$. \citeauthor{BMJ16} (\citeyear{BMJ16})  obtained $\tau_{r} = 20^{h}.9$ at 154 MHz. Conversion to 111 MHz leads to $\tau_{r} = 1^{d}.8$. Hence, the estimate of the scintillation index (RISS) from equation (3) is very close to the corresponding value at 74 MHz. However, values of $\tau_{r}$, obtained by equation (4) differ by more than 50 times. This problem has been discussed already by \citeauthor{GRC93} (\citeyear{GRC93}). They predicted $\tau_{r} =98^{d}$ using the equations valid for an extended medium model and a Kolmogorov spectrum without an inner scale (\citeauthor{RCB84}, \citeyear{RCB84}). This value differed by 30 times from the measured one $\tau_{r} =3^{d}.2$. The authors concluded that "It is not clear whether the results for PSR 0950+08 are part of the same trend for the predicted RISS time scales to overestimate the observed values, or whether it presents a special case". Thus, we have two time scales for RISS at 111 MHz. Both scales are seen in our data (Fig. 1).These scales are several days and several months. Perhaps we have deal indeed with a special case in the direction to this pulsar. Hence, the observed variations of flux densities in the pulsar B0950+08 at the scales of order of a few days and months can be explained by RISS.

Analyzing intensity variations we must take into account that the LPA radio telescope received radiation with a linear polarization. It is known that the position angle in the pulsar B0950+08 changes smoothly by $180^{\circ}$ between MP and IP (\citeauthor{BR80}, \citeyear{BR80}). High degree of polarization is observed at frequencies lower than 200 MHz For example, it is $67 \pm 6\,\%$ at the frequency of 39 MHz (\citeauthor{SVM83}, \citeyear{SVM83}). The rotation measure of this pulsar is $4\,\text{rad}\,m^{-2}$ and the period of the intensity modulation due to the Faraday rotation is 6 MHz at the frequency of 111 MHz (\citeauthor{SS04}, \citeyear{SS04}). Since the receiver bandwidth is 2.35\,MHz the angle of rotation is $70^{\circ}$. This leads to the noticeable and different intensity variations in two components of MP. (\citeauthor{SS04}, \citeyear{SS04}) showed that the ratio of flux densities of these components could be up two times depending on the degree of the linear polarization. Since the position angle between MP and IP changes approximately by $180^{\circ}$ the ratio of amplitudes of MP and IP must be unchanged for different observation days. However, intensities of the bridge and the precursor will change from day to day. We have analyzed the precursor (Fig. 4), taking into account its high intensity of $29\%$ comparing to MP (August 1, 2017, Table 1, Fig.4). The detailed description of errors of each considered component is given in Section 3.5. We have taken into account here  one linear polarization of our antenna. The small sample of days with strong emission of the bridge does not allow us to make the detailed analysis of intensity variations of this feature. It is known that MP in this pulsar contains the orthogonal component   manifested in the position angle (see, for example, \citeauthor{MHM80}, \citeyear{MHM80}). This can explain the rear cases when we see IP and Pr but there is no MP (pulses 179 and 741 in Fig. 2).Sometimes we observe only one component of MP, as a rule C2. Figure 2 shows such cases.

\subsection{Integrated profiles}

One of the main aims of our work was the detailed analysis of the pulsar emission for August 1, 2017. Let us compare the integrated profile of this day with other days. We constructed three groups dividing they by their intensities. Pulses in the first group have on average $<S/N>$ = 250, for the second one $<S/N>$ = 590, the third one is characterized by $<S/N>$ = 1100 (Table 1). Further we compare phases of MP and IP, their relative intensities, distances between MP and IP and their widths by the level 50\%.

\begin{figure}[ht]
\centering
	\includegraphics[width=\textwidth]{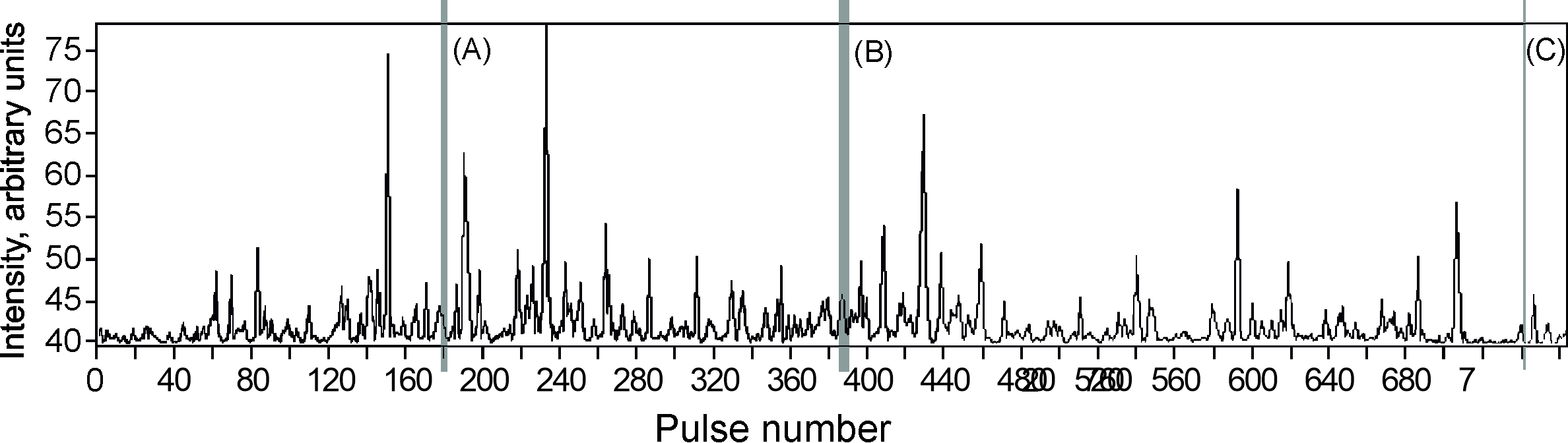}
    \caption{\small Relative intensity of radiation during the observation session on August 1, 2017. The grey blocks (A, B, C) are the pulses shown in Fig. 2.}
\end{figure}

\begin{figure}[ht]
\centering
	\includegraphics[width=\textwidth]{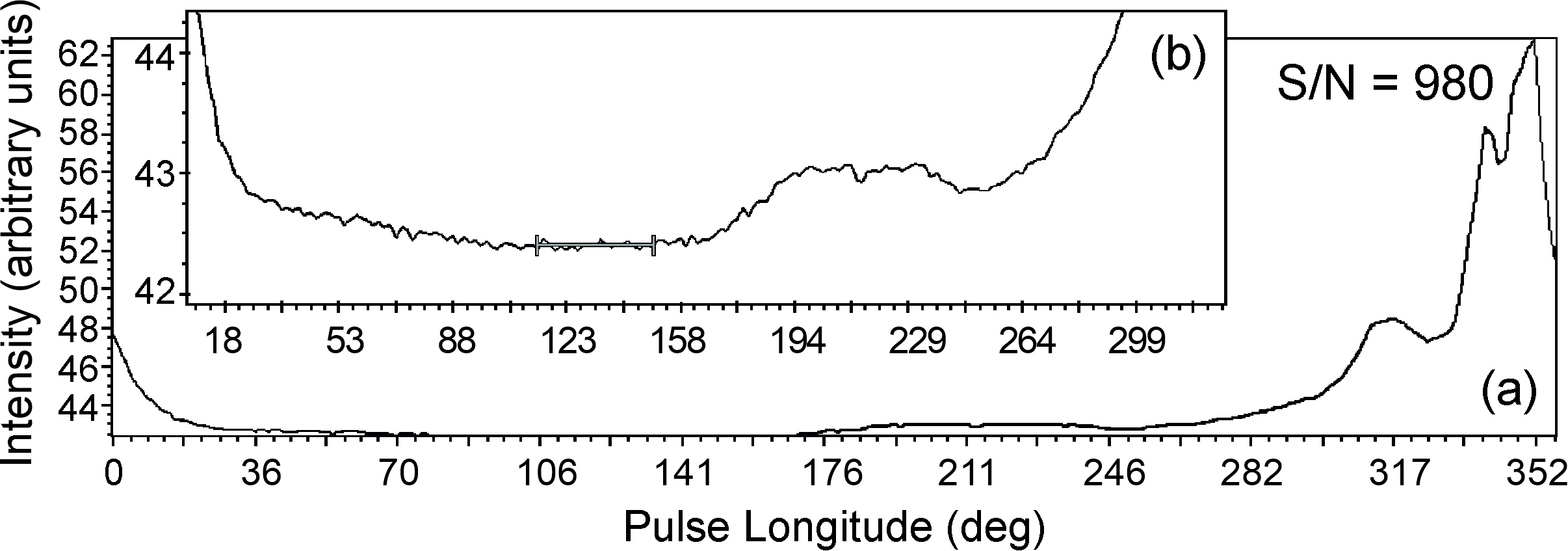}
    \caption{\small Integrated profile of the PSR B0950+08 on August 1, 2017 (a) and outside the main pulse,  85\%  of the whole period at a scale of 10\% of the amplitude of MP (b). The zero level is marked by the segment between $114^{\circ}$ and $150^{\circ}$ of the pulse longitude (b).}
\end{figure}   

Figure 4a shows the integrated profile for August 1, 2017 (sum of 764 pulses). This day was unique among all 450 observation days. The pulsar emitted   practically during all time (see, for example, some successive pulses in Fig. 2). \citeauthor{PL85} (\citeyear{PL85}) and  \citeauthor{HC81} (\citeyear{HC81}) gave the integrated profile  of the pulsar B0950+08 at the frequency of 408 MHz and declared that emission was observed almost during the entire period (83\%) and individual interpulses were occasionally as strong as 60\% of the average main pulse. It follows from Fig. 4b demonstrating 85\% of the whole period without the main pulse and drawn at a scale of 10\% of the amplitude of MP that the interpulse emission (IP and the bridge) is on average at least 2.5\% of the MP intensity and this radiation occupies about 90\% of the entire period. Fig. 4 shows clearly that the IP emission passes smoothly into the bridge and then through the precursor into MP. The similar effect was noted earlier by \citeauthor{SS88} (\citeyear{SS88}) at the frequency of 102.5 MHz, but their estimate was approximately two times lower (1.3\%). Since our estimate has been obtained on the whole period but not at the specific phase inside it we declare the much more intense effect. Moreover this estimate exceeds the mean level for all three groups mentioned earlier (Table 1). As can be seen from this table the increasing of the MP intensity correlates with the decreasing of the intensities for Pr and IP. Comparison of intensity on August 1, 2017 with other days shows that it is several times higher than S/N for the third most power group.

\begin{table}[ht]
	\centering
	\caption{Observed intensities for three groups mentioned.}

	\begin{tabular}{lccc} 
		\hline
		Data & S/N & Intensity of precursor & Relative intensity\\
		 &  & in relation to & of IP/MP, (\%)\\
		  & & the main pulse,(\%) & \\
		\hline
		07.10.2018 & 225 & 3.9 & $< 0.5$\\
		04.11.2018 & 297 & 7.0 & 1.0 \\
		03.01.2019 & 247 & 10.5 & 2.5 \\
		02.03.2019 & 211 & 5.0 & $< 0.5$\\
		\\
		11.09.2018 & 481 & 2.6 & 2.5 \\
		01.12.2018 & 568 & 12.3 & $< 0.5$ \\
		04.12.2018 & 571 & 5.3 & 2.5 \\
        05.04.2019 & 648 & 5.9 & 1.2 \\
        05.05.2019 & 688 & 6.3 & 1.2 \\
        10.05.2019 & 585 & 2.7 & 2.5 \\
        \\
        13.07.2017 & 1656 & 9.6 & 0.8 \\
        \textbf{01.08.2017} & \textbf{980} & \textbf{29.0} & \textbf{3.5} \\
        05.08.2017 & 1368 & 5.2 & 1.3 \\
        06.03.2018 & 984 & 8.2 & 0.8 \\
        14.03.2018 & 795 & 3.4 & 1.8 \\
        03.07.2018 & 982 & 2.8 & 1.0 \\
        07.09.2018 & 1159 & 3.2 & 0.7 \\
        13.09.2018 & 1276 & 11.1 & 0.8 \\
        12.02.2019 & 948 & 16.8 & $< 0.5$\\
        10.03.2019 & 1014 & 7.3 & 2.3 \\
		\hline
	\end{tabular}
\end{table}

It is worth noting that in the pulsar considered it is difficult to determine the level where there is no radiation (\citeauthor{HC81}, \citeyear{HC81}). We took for the zero level an area with minimum radiation intensity. The duration of zero longitudes is $36^{\circ}$ for all days. The zero level is shown in Fig. 4b.

\subsection{Individual pulses}

Strong jumps in intensity have been observed on August 1, 2017 both in the separate components of MP and in the interpulse space. Fig. 2 shows series of individual successive pulses for this day. The central part of the window contains IP.  This feature can be as strong as MP (Fig. 2(B), №389) and even stronger than MP (Fig. 2(C), №741).

As a rule the structure of pulses and their intensity changes sharply from one pulse to another (see Fig. 2(A), where four successive pulsar periods are illustrated). However, sometimes radiation is stable during 3-4 periods (Fig. 2(B)). As for the interpulse emission it changes markedly from one period to another. In Figure 2 grey blocks show the  phase boundaries of the components relative the integral form (Fig. 4). These are : IP = 175 - 230 deg, Pr = 300 - 323 deg, MP = 324 - 360 deg with the boundary between C1 and C2 at the phase of 342 deg.  All series of pulses are marked by the grey blocks at Figure 3.The interval of our observations (13.07.2017 - 10.03.19)  is inside of the interval of observations of this pulsar  using  LOFAR at the frequency of 60 MHz (\citeauthor{KWR20}, \citeyear{KWR20}). They showed that the flux density increased in 2018 comparing to 2017. The similar effect has been observed  at our frequency 111 MHz as well (see Fig. 1). Moreover, at 60 MHz  275 giant pulses (GPs) have been detected during  90 hours of observations, that is 3 GPs per hour. \citeauthor{TSA15} (\citeyear{TSA15}) have detected 5 GPs  during one hour observations at 39 MHz.  At our frequency about 250 strong pulses have been detected by \citeauthor{S12} (\citeyear{S12}) during one hour observation also.  However, it is not clear can we consider these pulses as GPs.  \citeauthor{SV12} (\citeyear{SV12}) declared that about  1 \% of pulses at 103 MHz are GPs. We found no one GP during 20 days (64 minutes)  of observations (Table 1). Such large discrepancy in the rate of GPs  at close frequencies can be explained  by the strong day-to-day variation of this rate (\citeauthor{SV12}, \citeyear{SV12}).  

As can be seen from Fig. 2 some individual pulses can give emission during the whole period of the pulsar. The maximal value of S/N for one of the individual MP is more than 2000. The mean value of S/N is 980. We carried out the quantitative analysis of the individual behavior of IP, Pr and MP for each of 20 days given in Table 1.

\subsection{Main pulse}

To find out how  the day of August 1, 2017 is unique we construct three groups of integrated pulses with different values of S/N. They are listed in Table 1.  The first group corresponds to 4 days of observations, the second one to 6 days and the third to 10 days. All 20 days are inside the interval between July 13, 2017 and May 5, 2019. First of all we counted numbers of individual pulses in 9 intervals of the signal to noise ratios from S/N $\geq 5$ to S/N $\geq 1280$, for each next interval S/N is two times higher. Then we analyzed dependences of these numbers on S/N in integrated pulses for all 9 intervals. Fig.5 shows the dependences for two intervals. It is obtained that for small values of S/N (Fig. 5a) the number of pulses decreases with the growth of S/N. For strong pulses (S/N = 320 - 640) the dependence is opposite. There is not a correlation for intermediate values of S/N.

\begin{figure}
	\centering
	\includegraphics[width=9cm]{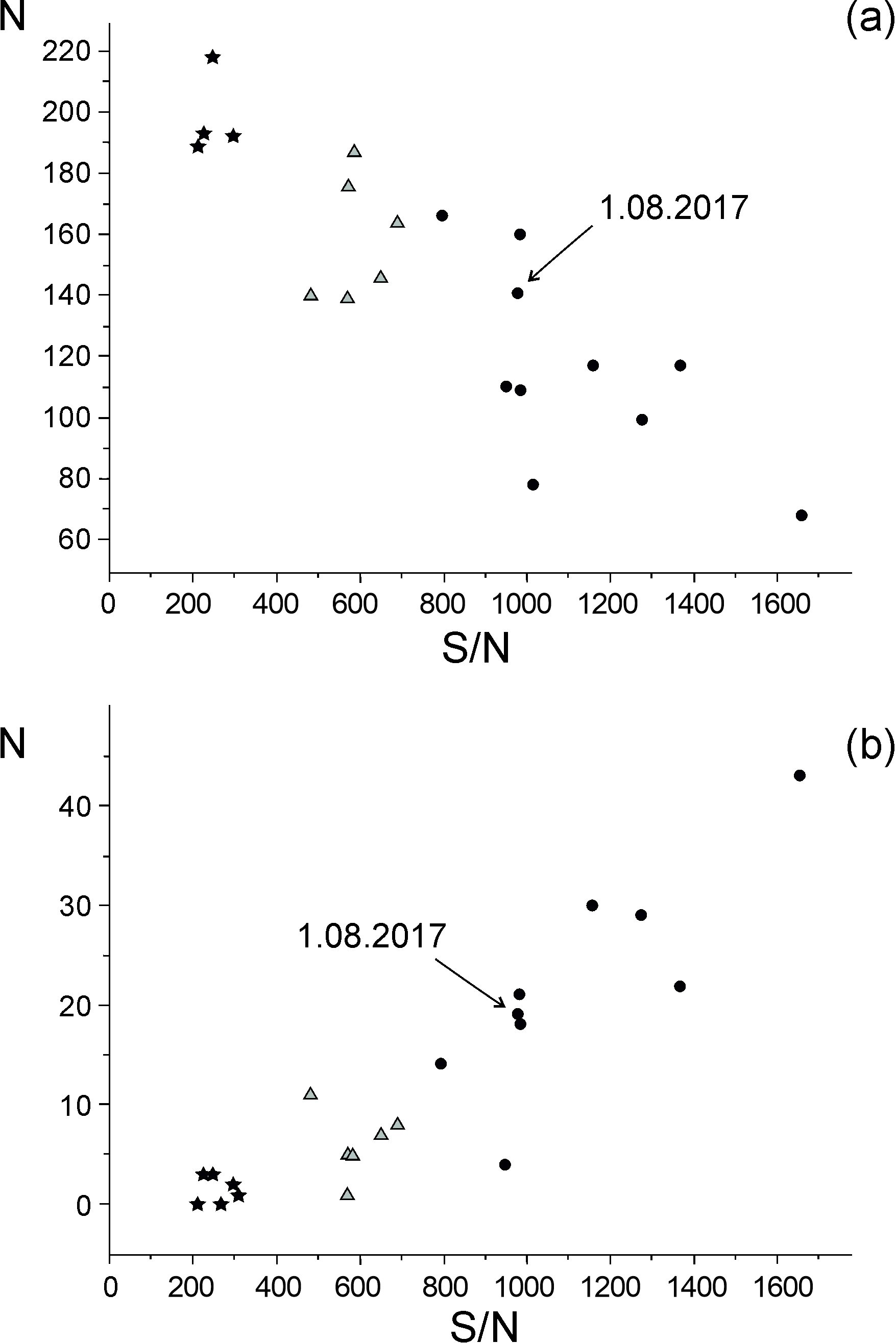}
    \caption{\small Numbers of individual pulses for S/N = 5-10 (a) and S/N = 320-640 (b).The mean S/N ratio is shown at the abscissa (see Table 1). Three groups of  days from Table 1 are shown by different marks: the first group by the asterisks, the second one by the grey triangles and the third one by the circles.}
\end{figure}
\begin{figure}[ht]
	\includegraphics[width=0.9\textwidth]{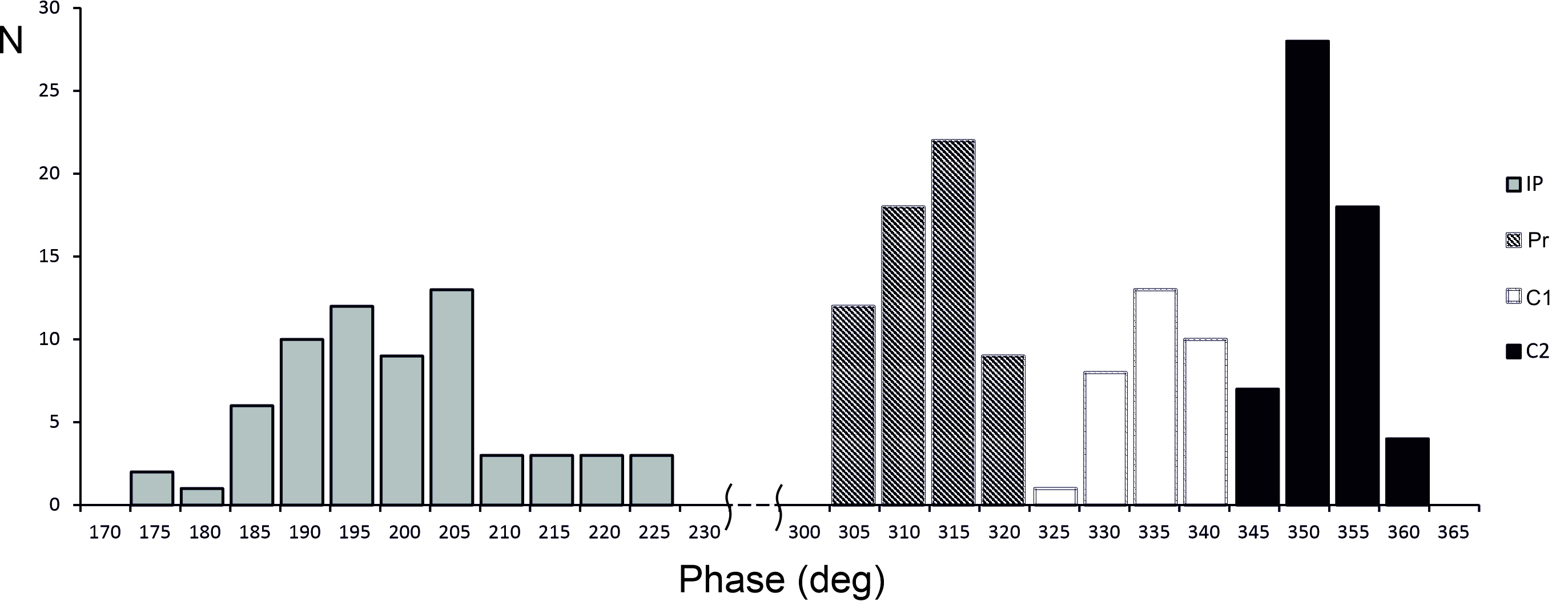}
    \caption{\small Distributions of phases for IP, Pr, C1 and C2.}
\end{figure}

We analyzed differential distributions of pulse numbers for each range of values of S/N. The maximum of the distribution for the first weak group corresponds to S/N $\geq 5-10$, and 66\% of pulses have amplitude with S/N $\geq 5$. In the third strong group the maximum of the distribution is located in the range S/N $\geq 10-40$, and the total fraction of visible pulses is of order of 90\%. Hence, an amplitude of an integrated pulse depends not only on a number of individual pulses with an amplitude greater than the specified one but on an intensity of these pulses.

\subsection{Interpulse, precursor and main pulse}

The unique day 01.08.2017 is located in the middle of the strong integrated pulses (the third group) in Figure 5. However, the maximal value of the IP and Pr amplitudes were recorded just on 01.08.2017 both in the integrated profile and in individual pulses. \citeauthor{S06} (\citeyear{S06}) analysed the precursor and MP at the frequency of 111 MHz and then searched for giant pulses in this pulsar. We give detailed analysis of IP and its connection with Pr and MP for the first time. We analyzed 4 components of individual pulses, namely, IP, Pr, C1 and C2 of MP. The following parameters of these components have been investigated: phases, amplitudes, durations at level 50\%, and integrated energies. Besides, we considered the energy of the previous MP, since \citeauthor{HC81} (\citeyear{HC81}) found the connection of this component with IP. These parameters are listed in Table 2. Most of our flux density measurements was made using the calibration discrete sources and the calibration signal, which was recorded synchronously in the beginning of every  period of the pulsar. Such method was used earlier for the measurements of flux densities for 235 pulsars at 102.5 MHz  (\citeauthor{MMS00}, \citeyear{MMS00}). In the present work we tried to investigate the pulsar profile during entire period without any calibration signal. Therefore, we used new  LPA pulsar flux calibration hardware and software, described in \citeauthor{TTO16} (\citeyear{TTO16}); \citeauthor{TKT20} (\citeyear{TKT20}). Daily measurements of the noise flux densities, using 6 calibration sources were carried out. The calibration signal was recorded in the form of OFF–ON–OFF. The temperatures of ON and OFF signals were $2400\,K$ and $\sim 300\,K$, respectively. Sigma of noises was calculated in the interval with minimum radiation in the pulsar period (Fig. 4b). We obtained  $\sigma = 1.3 \pm 0.2 Jy$, with  the parameters of observations: time resolution - 1.23 msec and the total bandwidth - 2.245 MHz.  The processing of data gave  S/N  for single pulses and we could calculate the energy   of  all components. In Table 2 all 65 individual pulses with the signal to noise ratio more than 4 are collected. The first column contains  the pulse numbers. From the second column to the fifth one we give phases of 4 components. The sixth and seventh ones present the phase differences D with their errors (in brackets). The eighth    column shows for which component value of D is calculated, when the amplitudes of components C1 and C2 not equal. For examples, there are pulses 386, 387 (Fig. 2b), where we can see component C2 only.  From the ninth column to the twelfth one energies are given. Fig. 6 shows the  distributions of phases for all four components. We can note two features in these distributions. Firstly, three components (Pr, C1 and C2) can be approximated by gaussians.  Secondly, the size of the distribution for IP is approximately equaled to the total size of the Pr+MP distribution. We search for connections between IP and other components (Pr, MP, and separately C1 and C2). More than 20 paired relationships between parameters of different components (phases, amplitudes, durations, and energies) were analyzed.  The strong inverse correlation between  the phase of IP and the difference D  of phases  (Pr-IP) has been detected (Fig. 7a).

The correlation coefficient for 62 pairs is K = - 0.92. The probability of the random distribution in Figure 7a $p < 0.0001$.  The linear approximation of the relationship in Figure 7a can be described by the following equation:
\begin{equation}
    D = (-1.02 \pm 0.06)\text{Phase(IP)}+(318 \pm 11),  
\end{equation}

There is also the inverse dependence between phases of IP and MP-IP (Fig. 7b). It shows the high correlation coefficient $K = - 0.85$ and the probability of the random distribution $p < 0.0001$.   It is very important to know which component connects with IP. We conclude that this is Pr. Firstly, we can see (Table 1, Fig. 4) that the  intensity of Pr in the integral pulse on 01.08.2017 is much higher than during other days and is about 29\% of MP. Secondly, when IP appears  Pr, as a rule, is observed  also (62 cases from 65 ones) (Table 2).  In the same time we see 57 C2 and 32 C1 only. Fig. 2 shows clearly that  all 7 pulses have IP and Pr simultaneously , and only 4 MPs in the same time. Third, there are positive correlations between energies of different components (Fig. 8). For 62 pairs of Pr and IP we obtained:
\begin{equation}
    \log E(\text{Pr}) = (0.48 \pm 0.11)\log E(\text{IP})+(1.49 \pm 0.24),
\end{equation}
with the correlation coefficient $K=0.49$ and the probability of the random distribution $p < 0.0001$. For  other two relationships we have for 60 pairs of MP and IP $K=0.17$, $p=0.18$, for 55 pairs of IP and the previous MP $K=0.26$, $p=0.05$.

We have calculated the similar relationships for 3 other days (11.09.2018, 10.03.2019 and 10.05.2019) with intensities of IP more than 2.3\%  comparing to MP (Table 1). The total number of Pr-IP pairs  was equal to 23. The mean value of the correlation coefficient  ($<K>=0.44$) is very close to the mentioned value of $K=0.49$ for 01.08.2017. For 53 IP-subsequent MP pairs $K=0.30$, and for 52 IP-previous MP pairs $K=0.36$. These values are somewhat higher than $K=0.26$ for 01.08.2017. It is important to note that obtained values of $K$ (0.26 and 0.36) for IP-previous MP pairs are close to the values from \citeauthor{HC81} (\citeyear{HC81}), where this quantity changes from 0.14 to 0.34 for 3 series counting 200 pulses.

\begin{figure}
\centering
	\includegraphics[scale=0.62]{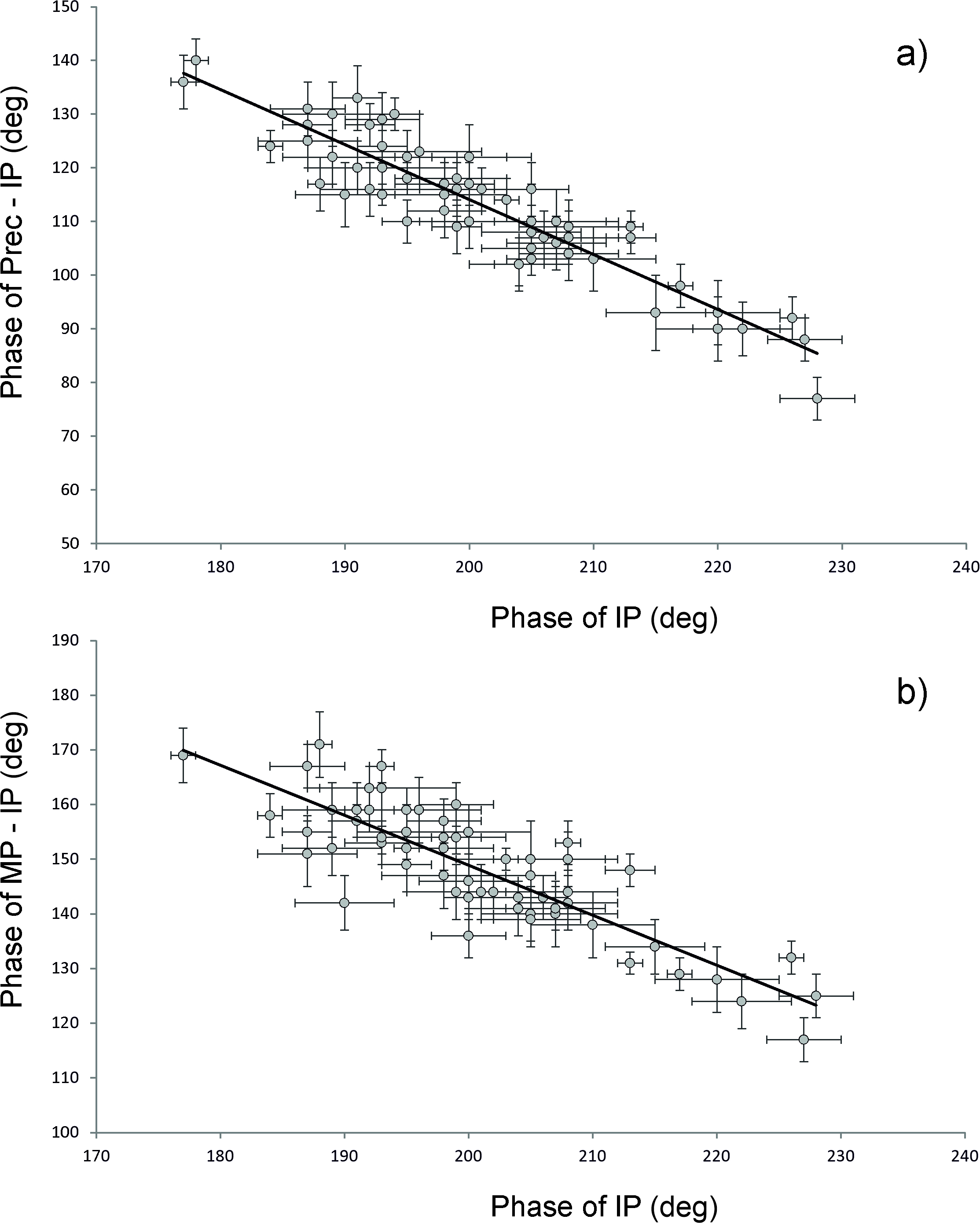}
    \caption{\small Relationship between phases of IP and Pr (a); between IP and MP (b). }
\end{figure}

Let us discuss the problem of energy measurement errors taking into account the registration of signals by the linearly polarized antenna As follows from Section  3.1 IP and MP are 150 degrees apart and have the same phase of the position angle (the difference is equal to 180 deg.). However, paying attention to the rather wide distributions of phases for different components of pulses (Fig. 6), we must take into account the corresponding changes of position angles  when estimate errors in values of energy. We have estimated the mean statistical errors  of the measured values of energy. These are for IP $\varepsilon=\frac{22^{\circ}}{75^{\circ}} \times 70\% =21\%$, for MP  $\varepsilon=\frac{25^{\circ}}{75^{\circ}}\times70\% =23\%$ . Here 22 deg and 25 deg are the widths of IP and MP distributions, and 75\% is the part of the period where the  changes  of  the position angle are equal to 90 deg. The  degree of the pulsar  linear polarization is 70\% at our frequency (see Section  3.1). The situation is not so clear for the IP-Pr pair. If we suggest that the Prs are on the average 112 deg. apart  from IPs (Fig. 6), then the error of the measured energy in Pr  consists of two quantities. The first one is caused by  the phase variations for Pr $\frac{12^{\circ}}{75^{\circ}} \times 70\% =  11\%$, the second by the difference of position angles of these two components  $\frac{112^{\circ}-75^{\circ}}{75^{\circ}} \times 70\% =35\%$.  These errors are independent and their mean is $37\% = (11^2+35^2)^{1/2}$. Also, there are errors of measurements of energy itself caused by noises. This one is no more than 15\%. Therefore the total errors in energy are 26\% for IP and 28\% for MP. The error is equal to 40\% for Pr. We don't know the position angle location for the IP-Pr pairs during our observations and have taken the mean value of the error for both components $\frac{26\%+40\%}{2} = 33\%$. We present 3 relationships  between energies of IP-Pr, IP-MP and IP-previous MP in Fig. 8.  We did not show errors. They  greatly burden the presented pictures, but don't change the obtained dependences  and values of the correlation coefficients.
 
 \begin{figure}
	\centering
	\includegraphics[width=9cm]{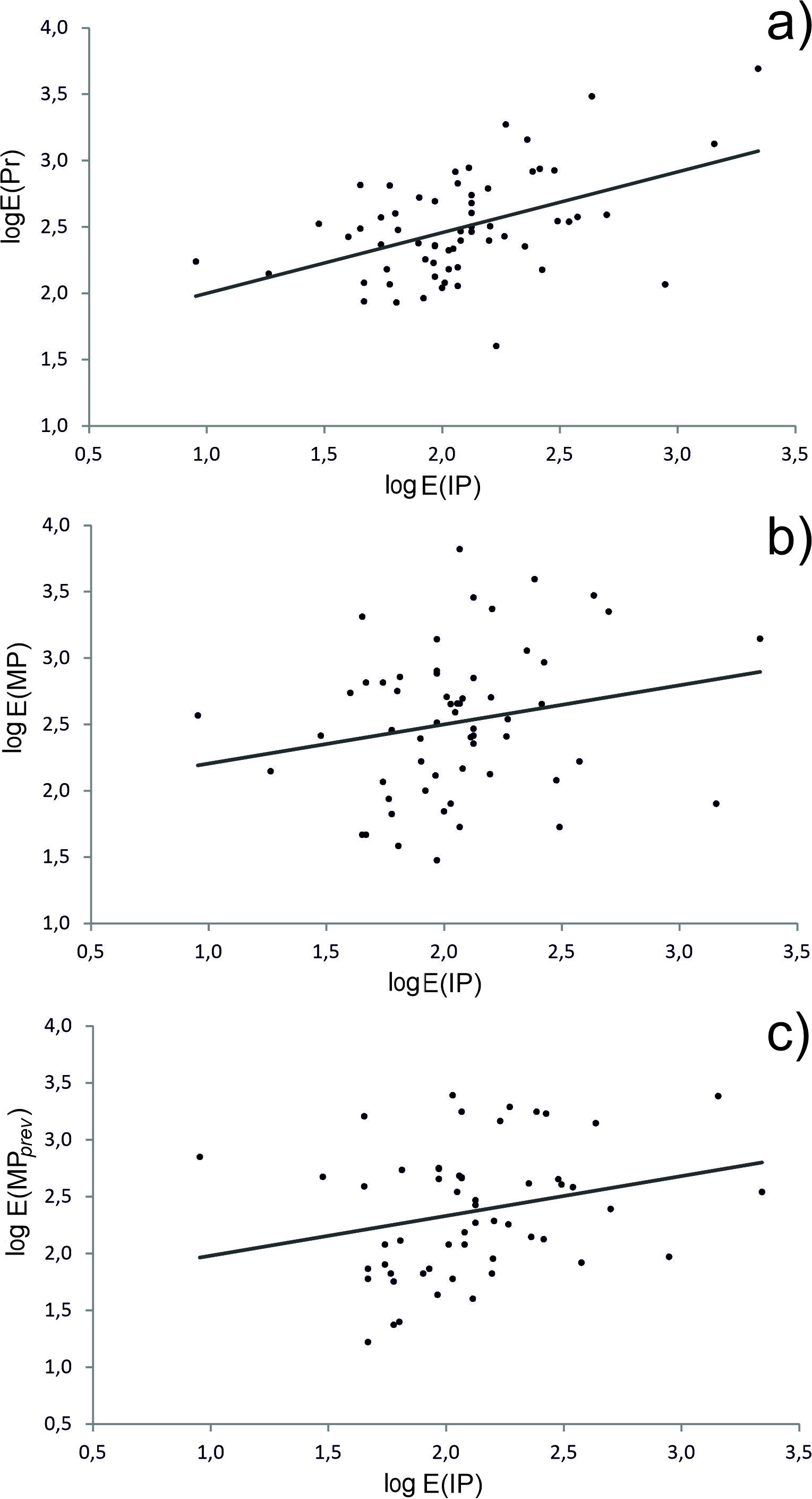}
    \caption{\small Relationship between energies of Pr and IP (a); MP and IP (b); previous MP and IP (c).}
\end{figure}

We can conclude that there are correlations both between intensities of IP and Pr and between their phases. This picture reminds the complex relation between the interpulse and the second component of the main pulse in the pulsar B1822-09 (\citeauthor{GJK94}, \citeyear{GJK94}) where the correlations between spectra and durations of these features have been detected.

\section{Discussion}

\citeauthor{HC81} (\citeyear{HC81}) have given the following arguments for the assumption on the pulsar B0950+08 as an aligned rotator. 1) There is emission between MP and IP. 2) There is the correlation between intensities of MP and IP. 3) The distance between MP and IP differs from $180^{\circ}$. 4) The position angle of linear polarization changes smoothly from MP to IP. 5) The microstructures in MP and IP have the similar time scales: $130\,\mu$sec in MP and $90\,\mu$sec in IP (\citeauthor{BBM73}, \citeyear{BBM73}). We believe that these arguments exclude the possibility of a description of observed peculiarities in the frame of the orthogonal rotator model. \citeauthor{MN13} (\citeyear{MN13}) estimated values of the angle $\beta$ between the magnetic moment and the rotation axis in the pulsar B0950+08 by several methods and gave the mean value of this angle $<\beta> = 18.9^{\circ}$. For the angle between the rotation axis and the line of sight they obtained $\zeta = 6.7^{\circ}$.  For these angles the model path of the position angle of linear polarization is in a good agreement with the observed one. Earlier the similar values of these angles have been obtained by \citeauthor{NV83} (\citeyear{NV83}), namely $\beta = 10^{\circ}$ and $\zeta = 5^{\circ}$. Indirect argument for the aligned rotator in the case of PSR B0950+08 is its age. \citeauthor{MN13} (\citeyear{MN13}) showed that orthogonal rotators had ages several times less than aligned ones. The investigated pulsar locates far enough from the Galactic plane, its height is $z = 0,18 kpc$.  The low radio luminosity of this pulsar ($L = 27.41 mJy\cdot kpc^{2}$) shows also that it is the decaying source. Moreover using its transverse velocity $V = 36.55 km/sec$ (\citeauthor{MHT05}, \citeyear{MHT05}) and suggesting that its moving is isotropic, i.e. $V_{z} = 25.84\,km/sec$, we can estimate its kinematic age as $\tau=z/V_{z}=6.8$ millions of years. This means that it is the old pulsar. Suggesting that the braking of the pulsar provided with the constant rate $dP/dt = 2.3\cdot10^{-16}$ (\citeauthor{MHT05}, \citeyear{MHT05}) and its real age equals to the kinematic one we can estimate its initial period obtained in the moment of its birth $P_{0} = P - \tau dP/dt \approx 0.2\,sec$. This means that the pulsar B0950+08 was born as a long periodic object, and that not all pulsars were born with millisecond periods. If PSR B0950+08 is a typical object among pulsars we can conclude that old pulsars have small values of the angle $\beta$ and, hence, these angles decrease with the age. Such a conclusion was made earlier by \citeauthor{M90} (\citeyear{M90}) and by \citeauthor{PTL14} (\citeyear{PTL14}).

\begin{figure}[p]
	\centering
	\includegraphics[scale=0.7]{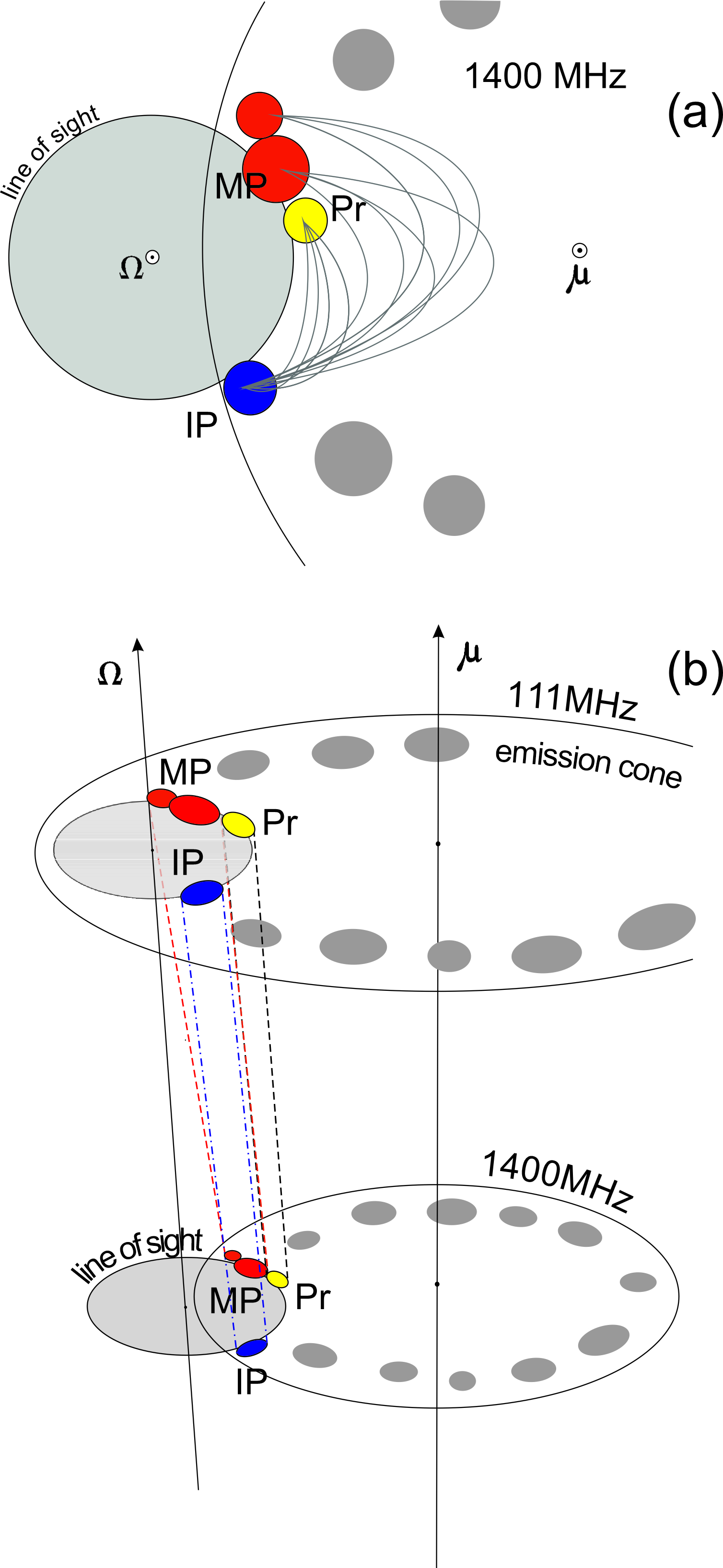}
    \caption{\small Proposed explaining of the obtained dependences between parameters of MP and IP at two frequencies in the frame of the single pole model. a) Relative distribution of spots at the surface of the neutron star, rotation ($\Omega$), magnetic axes ($\mu$) and line of sight at the level where radiation at the frequency of 1400 MHz is generated are shown. Axes $\Omega$ and $\mu$ are perpendicular to the plane of the Figure. b) MP and IP emission is generated in the cones connected with some spots. Here it was assumed that the angle between the magnetic moment and the rotation axis $\beta = 18.9^{\circ}$ and between the line of sight and the rotation axis $\zeta = 6.7^{\circ}$. For the clarity the top cone angular  size is shown not in the same  scale.}
\end{figure}

The possible geometry explaining the peculiarities of observed emission is shown in Figure 9. We use further values $\beta = 18.9^{\circ}$ and $\zeta = 6.7^{\circ}$. Some active areas (spots) can exist in a polar cap of the pulsar. They are formed as a result of the surface sparking causing the generation of unsteady electric field (\citeauthor{RS75}, \citeyear{RS75}).Their radiation then spreads through the narrow columns with angular sizes increasing away from the surface of the neutron star. It is variable due to changes with time of concentrations and energies of emitting charges. Such a variability reveals itself in sharp changes of all observed components (MP, IP, Pr and emission between them). At low frequencies the main pulse consists of two components, at enough high ones one of them is out of sight. The angular distance between MP and IP can remain unchanged if the location of the spots mentioned is not changed.

We suggest here that the generation of emission at the chosen distance from the center of the neutron star occurs at the local plasma frequency
\begin{equation}
    \nu_{p}=\left(\frac{2n_{p}e^{2}}{\pi m}\right)^{1/2},
\end{equation}
where $n_{p}$ is the concentration of the secondary electron-positron plasma. Let us suggest that during the cascade process of the plasma birth all energy of the primary beams transferred to new electrons and positrons:
\begin{equation}
    \gamma_{b}n_{b}mc^{2}=2\gamma_{p}n_{p}mc^{2}
\end{equation}
Here $\gamma_{b}$ and $\gamma_{p}$ are Lorentz-factors of the primary particles and the new born electrons, respectively, and
\begin{equation}
   n_{b}=\frac{B}{Pce}
\end{equation}
is the concentration of charges in the primary beam (\citeauthor{GJ69}, \citeyear{GJ69}).

For the dipolar magnetic field we have
\begin{equation}
   \nu_{p}=2.37\cdot10^{3}\left(\frac{B\gamma_{b}R_{*}^{3}}{P\gamma_{p}r^{3}}\right)^{1/2}
\end{equation}

Substitution of $P = 0.253$ sec and magnetic  field at the magnetic pole, two times higher than the value of $B = 2.44\cdot10^{11}G$  from the ATNF catalogue (\citeauthor{MHT05}, \citeyear{MHT05}) and using values of $\gamma_{b} = 10^{7}$  and $\gamma_{p} =100$  give
\begin{equation}
   \nu_{p}=10^{12}\left(\frac{R_{*}}{r}\right)^{3/2}
\end{equation}

We calculated distances where emissions at frequencies 1400 and 111 MHz are generated: $\left( \dfrac{r}{R_{*}}\right)_{1400} = 90$,  $\left(\dfrac{r}{R_{*}}\right)_{111} = 489.$

The light cylinder radius $r_{\text{\textit{\tiny{LC}}}}=\dfrac{cP}{2\pi}$ is $1.2\cdot10^{9}\,cm = 1200\,R_{*}$ for the pulsar B0950+08. The equation of the last open field lines for the dipole geometry is
\begin{equation}
   \frac{r}{sin^{2}\theta}=r_{\text{\textit{\tiny{LC}}}}
\end{equation}

They form the emission cone in the frame of the polar cap model. Equation (12) determines the angular radii of this cone at the calculated levels: $\theta_{1400}=15.9^{\circ}$, $\theta_{111}=39.7^{\circ}$. These values have been used in Figure 9.

Our observations show the noticeable variations of the distance between IP and MP (Fig. 10).

\begin{figure}
	\centering
	\includegraphics[width=8cm]{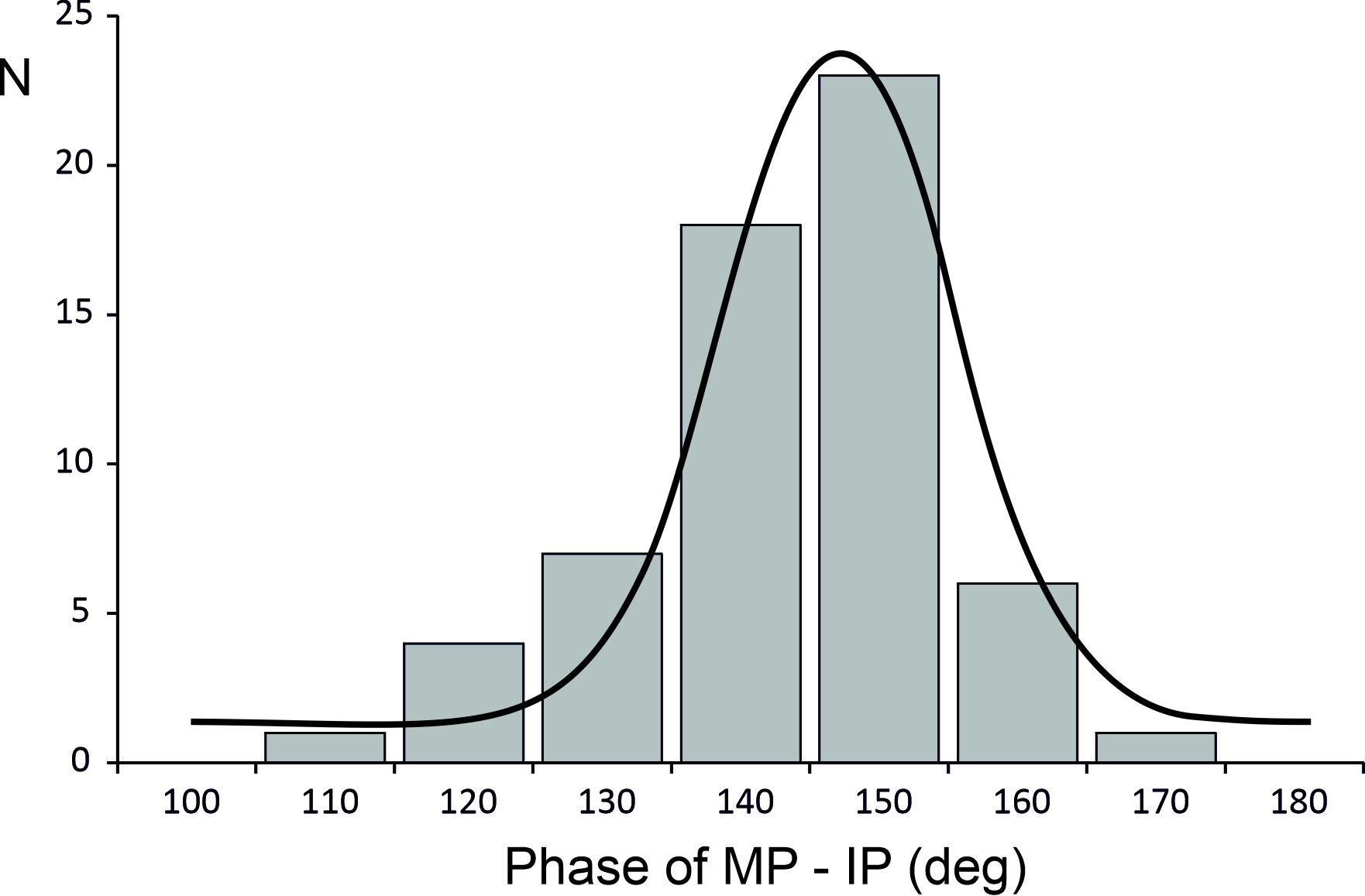}
    \caption{\small Distribution of observed distances between IP and MP.}
\end{figure}

The maximum of the distribution of D = IP-MP is $146.5^{\circ}$, and this distribution can be well described by the following Gaussian:

\begin{equation}
   N=24exp\left\{-\left(\frac{146.5-D}{16.4}\right)^{2}\right\}
\end{equation}

Variations of D, the wide distribution of this distance and the mentioned difference from $180^{\circ}$ are the strong additional arguments in favor of the alignment of this pulsar. These are mentioned here for the first time. For an orthogonal rotator D must be very near $180^{\circ}$ with minor deviations from this value. Variability in aligned objects can be caused by different intensities of separate parts of corresponding spots and by their movement on the surface of the neutron star caused by oscillations of the surface and waves spreading over it. Figure 9 shows how we can reconcile our model with the observed dependences of features in MP and IP on frequency given by \citeauthor{HC81} (\citeyear{HC81}).

In our model distances between sub-pulses of MP are determined by the structure of magnetic field lines in spots (Fig. 9a). At high frequencies near the surface of the neutron star this distance is almost constant. When we move away from the surface lines of force diverge, the angular radius of the emission cone increases and the distance between sub-pulses increases also. This effect explains similarly the width of MP at different frequencies. At a some distance from the neutron star one of the spots is out of sight and one of the components becomes invisible. The width of MP changes abruptly at higher frequencies and is determined by the width of the reminder component. The distance between IP and MP as is declared earlier is determined by the almost unchanged distance between the corresponding spots. They are connected by magnetic field lines (see Fig. 9), and this explains the weak radiation at longitudes between MP and IP.

\section{Conclusions}

Some results of long-term observations of the pulsar B0950+08 using radio telescope LPA at the frequency of 111 MHz are given. 

A strong variability of the intensity was revealed in radiation of the pulsar showing the changes of signal to noise ratio hundreds times at different time scales. This variability is caused mainly by peculiarities of the emission mechanism. However, long-term fluctuations with the characteristic time of several days and months can be explained by RISS in the interstellar medium.

The analysis of the integrated profile of MP shows that the amplitude of its radiation depends on both number of pulses and their intensities. The existence of radiation between IP and MP has been confirmed. It is more intense comparing to the higher frequencies. We carried out the detailed analysis of the unique event on 1.08.2017 showing strong radiation in the almost entire interpulse space. 

We detected the high correlations both between changes of intensities of Pr and IP and between phases of IP and Pr. New peculiarities of the pulsar emission such as the correlation between parameters of IP and Pr, the rather  wide distribution of IP phases and distances between IP and MP can be explained  in the frame of the  aligned rotator model. The empirical version of such a model of the pulsar magnetosphere is proposed. We believe that there are some spots at the surface of the central neutron star in the pulsar B0950+08. They determine the structure of emission cones at different frequencies. Particularly, such a model explains qualitatively the complex picture of emission at low frequencies and gives the possibility to calculate distances where the observed emission is generated.

We estimated also the initial period of the pulsar acquired at birth ($P_{0} \approx 0.2 s$). This means that not all pulsars are born with millisecond periods. The large age of the pulsar (6.8 millions of years) and the small angle between the magnetic moment and the rotation axis are the evidences of the evolution of pulsars to aligned rotators.\footnote{When our manuscript was ready for the submission the preprint by Bilous et al. (astro-ph HE 2109.08500) appeared. There some similar problems concerning variations of radiation of the pulsar B0950+08 at other frequencies (55 MHz and 1.4 GHz) are discussed. The detection of the rather strong feature between MP and IP described in this preprint is the additional argument for the aligned rotator model.}

\section*{Acknowledgements}

The authors thank T.V. Smirnova for the useful discussions and S.V. Logvinenko for the technical support of observations and data processing.


\begin{table}
	\caption{Characteristics of some components of individual pulses.}
	\begin{tabular}{cccccccccccc} 
	\hline
        Pulse  &  \multicolumn{4}{c}{Phase (deg)} &  \multicolumn{2}{c}{D (deg)} & Component & \multicolumn{4}{c}{Energy ($Jy\cdot msec$)} \\
         number  & IP & Pr & C1 & C2 & Pr - IP & MP - IP & of MP & IP & Pr & MP & MP$_{prev}$\\
        \hline
        85  & 199 & 317 & 343 &     & 118(2) & 144(5) & C1 & 92 & 169 & 130 & 43\\
        93  & 215 & & & & & & \\	
        109	& 191 & 311 &     & 350	& 120(2) & 159(5) & C2 & 184 & 268 & 256 & 180 \\
        128	& 198 & 313	& 338 & 352 & 115(6) & 154(6) & C2 & 107 & 152 & 450 & 2454\\
        145	& 206 & 313 & 338 & 360	& 107(4) & 143(4) &    & 79  & 238 & 246 & \\
        146	& 198 &	313 & 338 &	352	& 115(1) & 147(6) &    & 500 & 390 & 2231& 246\\
        177 & 195 &	317	&     & 350 & 122(2) & 155(5) & C2 & 346 & 346 & 253 & 383\\
        178	& 188 &	305	&     & 359	& 117(4) & 171(6) & C2 & 114 & 823 & 453 & 253\\
        179	& 178 & 318 &     &	    & 140(3) &        &	   & 113 & 478 & 293 & 453\\
        191	& 191 & 324	&     & 348	& 133(5) & 157(3) & C2 & 117 & 673 & 6593& 466\\
        211	& 205 & 315	&     & 355	& 110(3) & 150(7) & C2 & 376 & 376 & 167 & 83\\
        214	& 196 & 319	&     & 355 & 123(4) & 159(6) & C2 & 120 & 250 & 496 & 153\\
        224	& 207 & 317	&     & 347	& 110(3) & 140(6) & C2 & 266 & 150 & 929 & 1692\\
        228	& 189 & 319	& 340 & 348	& 130(4) & 159(5) & C2 & 133 & 316 & 2850& 186\\
        232	& 205 & 321 & 336 & 352	& 116(3) & 139(5) &    & 242 & 827 & 3926& 1765\\
        248	& 195 & 305 & 336 & 352	& 110(3) & 149(5) &    & 103 & 120 & 509 & 120\\
        250	& 217 & 315	& 340 & 352	& 98(3)	 & 129(3) &    & 30  & 333 & 260 & 473\\
        271	& 204 & 306	&     & 347	& 102(3) & 143(3) & C2 & 58  & 152 & 87  & 67\\
        273	& 198 & 315	&     & 350	& 117(3) & 152(5) & C2 & 225 & 225 & 1136& 413\\
        300	& 220 &	313	&     &     & 93(2)	 &        &    & 170 & 40  &     & 1462\\
        301	& 187 & 318 &     & 354	& 131(3) & 167(4) & C2 & 83  & 92  & 100 & \\
        341	& 200 & 317 & 343 &     & 117(3) & 143(4) & C1 & 47  & 87  & 47  & 17\\
        346	& 193 & 308 &     & 360	& 115(1) & 167(3) & C2 & 85  & 180 & 40  & 73\\
        374	& 208 & 315 &     & 352 & 107(3) & 144(4) & C2 & 133 & 549 & 226 & \\
        379	& 177 & 313	& 334 & 357	& 136(4) & 169(5) &    & 65  & 300 & 719 & 541\\
        386	& 199 & 308 &     & 353	& 109(4) & 154(3) & C2 & 157 & 616 & 133 & 67\\
        387	& 207 & 313 &     & 348	& 106(3) & 141(4) & C2 & 260 & 866 & 450 & 133\\
        388	& 193 & 313 &     & 356	& 120(2) & 163(7) & C2 & 300 & 842 & 120 & 450\\
        389	& 195 &	317	&     & 354 & 122(4) & 159(4) & C2 & 55  & 233 & 117 & 120\\
        391	& 215 & 308 & 343 & 354	& 93(5)  & 134(5) &    & 63  & 400 & 563 & 25\\
        392	& 187 & 312	& 333 & 354	& 125(3) & 151(6) &    & 93  & 493 & 799 & 563\\
        394	& 208 & 317 &     & 358	& 109(2) & 150(5) & C2 & 133 & 290 & 260 & 266\\
        408	& 220 & 310	& 331 & 348	& 90(1)  & 128(6) & C2 & 45  & 653 & 2045& 390\\
        416	& 228 & 305	&     & 353 & 77(2)  & 125(4) & C2 & 107 & 210 & 80  & 60\\
        428	& 200 & 310	& 340 & 352 & 110(2) & 146(5) &    & 186 & 1865& 346 & 1941\\
        429	& 222 & 312	& 338 & 354	& 90(3)  & 124(5) &    & 2198& 4928& 1399& 3468\\
        430	& 200 & 322 & 336 & 350 & 122(5) & 136(4) &    & 433 & 3040& 2957& 1399\\
        431	& 208 & 312 &     & 350 & 104(2) & 142(5) & C2 & 1432& 1332& 80  & 2957\\
        432	& 226 & 318 &     & 358 & 92(3)  & 132(3) & C2 & 55  & 373 & 653 & 80\\
        438	& 210 & 313 & 338 & 357 & 103(2) & 138(6) &    & 133 & 403 & 706 & \\
        441	& 208 & 315 &     & 361 & 107(2) & 153(4) & C2 & 112 & 216 & 390 & 346\\
        442	& 189 & 311 & 333 & 357 & 122(2) & 152(5) &    & 310 & 350 & 53  & 390\\
        444	& 205 & 308 & 	  & 352 & 103(2) & 147(3) & C2 & 60  & 646 & 286 & 57\\
         \hline
	\end{tabular}
\end{table}

\newpage
\begin{table*}
Table 2. (Continued)
\centering
	\begin{tabular}{cccccccccccc} 
	\hline
        Pulse  &  \multicolumn{4}{c}{Phase (deg)} &  \multicolumn{2}{c}{D (deg)} & Component & \multicolumn{4}{c}{Energy ($Jy\cdot msec$)} \\
         number  & IP & Pr & C1 & C2 & Pr - IP & MP - IP & of MP & IP & Pr & MP & MP$_{prev}$\\
        \hline
        447	& 201 & 317	& 338 & 352 & 116(3) & 144(5) &    & 160 & 320 & 2338& 193\\
        457	& 199 &	315 &     & 359	& 116(3) & 160(4) & C2 & 230 & 1439& 67  & 140\\
        473	& 200 & 322 & 	  & 355 & 122(2) & 155(5) & C2 & 93  & 230 & 30  & \\
        483	& 184 & 308 & 329 & 355 & 124(2) & 158(4) &    & 93  & 133 & 326 & 453\\
        494	& 192 & 320 & 342 & 359	& 128(3) & 159(4) &    & 93  & 226 & 763 & \\
        497	& 198 & 310	&     & 355 & 112(4) & 157(4) & C2 & 120 & 293 & 147 & 120\\
        508	& 193 & 322 & 337 & 355 & 129(4) & 153(4) &    & 64  & 85  & 38  & 130\\
        533	& 205 & 310 & 334 & 355 & 105(3) & 140(5) &	   & 100 & 110 & 70  & \\
        549	& 192 & 308	&     & 355 & 116(3) & 163(4) & C2 & 130 & 882 & 253 & 40 \\
        579	& 227 & 315 & 336 & 352 & 88(2)  & 117(4) &    & 158 & 250 & 504 & 90\\	
        601	& 190 &	305	& 332 &     & 115(4) & 142(5) & C1 & 117 & 157 & 53  & 1758\\
        621	& 204 & 306 & 339 & 350 & 102(3) & 141(5) &    & 9   & 173 & 370 & 706\\
        647	& 213 & 322 & 334 & 354	& 109(2) & 131(2) &    & 117 & 113 & 455 & 460\\
        658	& 213 & 320 &     & 361	& 107(1) & 148(3) & C2 & 60  & 117 & 67  & 24\\
        667 & 193 & 317 & 341 & 353 & 124(2) & 154(3) &    & 47  & 120 & 653 & 60\\
        669	& 203 & 317 &     & 353 & 114(3) & 150(2) & C2 & 45  & 306 & 47  & 1605\\
        674	& 202 &	    & 341 & 350	&        & 144(3) &    & 93  &     & 1382& 554\\
        678	& 195 & 313 & 340 & 355 & 118(2) & 152(3) &    & 18  & 140 & 140 & \\
        704	& 206 &     &     &     &        &        &    & 47  &     &     & 73\\
        705	& 205 & 313	& 339 & 349 & 108(2) & 139(5) &    & 40  & 266 & 546 & \\
        741	& 194 & 324 &     &     &        & 130(2) &    & 886 & 117 &     & 93\\
        754	& 187 & 315 & 331 & 354 & 128(2) & 155(3) &    & 80  & 526 & 167 & 67\\
        \hline
	\end{tabular}
\end{table*}


\end{document}